\documentclass{optica-article}

\journal{opticajournal} % for journals or Optica Open

\articletype{Research Article}
\usepackage{float}
\usepackage{subfigure}
\usepackage{xcolor}
\usepackage{graphicx}% Include figure files
\usepackage{dcolumn}% Align table columns on decimal point
\usepackage{bm}% bold math
\usepackage{physics}
\usepackage{siunitx}
\usepackage{ulem}
\usepackage{mathrsfs}
\usepackage{lineno}
\usepackage{hyperref}% add hypertext capabilities
\hypersetup{
  colorlinks   = true, %Colours links instead of ugly boxes
  urlcolor     = blue, %Colour for external hyperlinks
  linkcolor    = red, %Colour of internal links
  citecolor   = red %Colour of citations
}

\renewcommand{\vec}[1]{\mathbf{#1}}

\begin{document}

\title{Wavefront correction of high-dimensional two-photon states via coherence-entanglement transfer}
% Maybe soemthing ont he line: Protecting high-dimensional entanglement via wavefront shaping and Schmidt number control. 

\author{Shaurya Aarav, Hugo Defienne\authormark{*}}

\address{Sorbonne Universit\'e, CNRS, Institut des NanoSciences de Paris, INSP, F-75005 Paris, France}

\email{\authormark{*}hugo.defienne@insp.jussieu.fr} %% email address is required; see note below about the corresponding author designation

% use {asbstract*} to suppress the copyright line. Copyright information will be added in production

\begin{abstract*} 
Reliable transmission of quantum optical states through real-world environments is key for quantum communication and imaging. Yet, aberrations and scattering in the propagation path can scramble the transmitted signal and hinder its use. A typical strategy is to employ a classical beacon beam to learn and then correct for the wavefront distortions. However, relying on a separate light source increases the overhead in the experimental apparatus. Moreover, the beacon light must closely match the non-classical state in polarization, wavelength, and even temporal bandwidth, which is highly challenging in practice. Here, we introduce a fast and efficient wavefront correction approach where we use the quantum state itself to correct for optical distortion. Via pump shaping, we control the degree of entanglement in the spatially-entangled two-photon state so that it behaves either as a high-dimensional entangled state or as a classical coherent state. The latter case is used to efficiently measure the transmission matrix of the propagation channel and correct its distortions with a spatial light modulator, thereby enabling the transmission of the high-dimensional entangled state with minimal errors. 
Our approach paves the way for the practical implementation of quantum imaging and communication protocols based on high-dimensional spatially entangled states.
%For this, we use a spatial light modulator to both correct for the aberration and shape the photon correlations. 
\end{abstract*}

%%%%%%%%%%%%%%%%%%%%%%%%%%  body  %%%%%%%%%%%%%%%%%%%%%%%%%%
\section{Introduction}
With the increasing maturity of quantum optical technologies, communication and imaging are poised to undergo a radical transformation. In this context, quantum states entangled in high dimensions~\cite{erhard_advances_2020}, such as those involving the transverse spatial degree of freedom~\cite{walborn_transverse_2007}, are particularly promising. They enable novel imaging platforms~\cite{defienne_advances_2024}, including those enhancing signal-to-noise ratio~\cite{brida2010experimental}, image contrast~\cite{defienne_quantum_2019,gregory_noise_2021,camphausen_quantum-enhanced_2021}, spatial resolution~\cite{toninelli_resolution-enhanced_2019,defienne2022pixel,he_quantum_2023} and removing aberrations~\cite{cameron2024adaptive}. In quantum communication, high-dimensional states offer increased information capacity and greater resilience to noise and losses~\cite{srivastav_quick_2022, ecker2019overcoming, zhu_is_2021,cozzolino_high-dimensional_2019}, compared to qubit-based approaches.

However, these states are sensitive to any spatial inhomogeneity, either in the propagation channel for communication, such as air turbulence, or when interacting with the object's environment, such as body tissue in imaging applications. Regardless of the origin, these effects potentially degrade the information transmitted via quantum correlations. Correcting these aberrations is therefore essential for restoring the system’s performance.

A common approach to this problem involves (i) using a bright coherent laser beam as a beacon to measure optical aberrations, (ii) programming a spatial light modulator (SLM) accordingly to compensate for them, and (iii) substituting the classical source by the photon-pair source so that it benefits from the correction. Using this method, two-photon states have been successfully controlled and transmitted through complex media - including scattering layers and multimode fibers - via optimization and transmission matrix (TM) based techniques~\cite{defienne_two-photon_2016,defienne2018adaptive,courme_manipulation_2023,devaux_restoring_2022}. 
In these works, the classical beacon was either an external laser at the photon-pair wavelength or directly the pump laser of the nonlinear crystal generating the pairs.
These approaches benefit from the broad and well-established toolbox of classical wavefront shaping algorithms~\cite{cao_shaping_2022} while circumventing practical challenges inherent to photon-pair sources, such as low intensity and the difficulty of coincidence detection.

Yet, using a classical beacon in quantum systems has limitations. First, it is an overhead component requiring extra space and alignment considerations when it is not useful for the intended application of communication or imaging. Second, and more importantly, the optical aberrations and the associated wavefront corrections are sensitive to several degrees of freedom, such as wavelength~\cite{xu2018imaging,mounaix_spatiotemporal_2016}, polarization~\cite{mounaix_control_2019, aguiar_polarization_2017}, temporal bandwidth~\cite{katz_focusing_2011,paudel2013focusing,mounaix_deterministic_2016} and angle of propagation~\cite{katz_looking_2012}. Thus, for the correction to be effective, the classical beacon must closely match the mode occupied by the quantum state in all the relevant degrees of freedom - a requirement that becomes more challenging as the complexity of the quantum state increases. 
For example, if the two-photon state is highly multimode, or if the photons have different wavelengths - as in ghost imaging~\cite{aspden_photon-sparse_2015} or imaging with undetected photons~\cite{lemos_quantum_2022} - it becomes necessary to carefully align two different lasers onto specific spatial modes, allowing only partial correction of the aberrations.
This motivates the development of fully quantum wavefront correction methods, where the same quantum light used for information transfer is also used to probe and correct optical aberrations and scattering.

\begin{figure*}[t]
%\centering
\includegraphics[keepaspectratio=true, scale = .74,trim = {2.cm 23cm 1.7cm 1.5cm},clip]{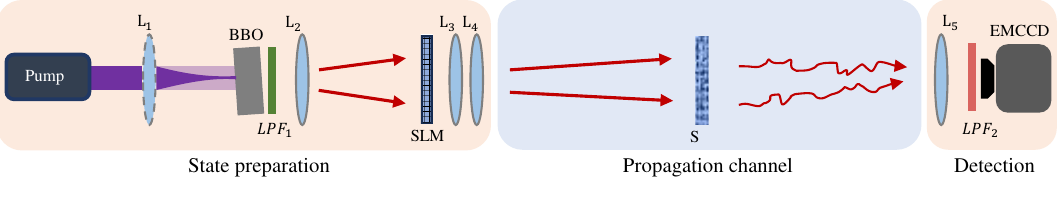}% Here is how to import EPS art
\caption{\label{fig_setup} \textbf{Experimental setup}. The setup involves three parts. The first is the state preparation, where a spatially-entangled two-photon state is produced and shaped with a spatial light modulator (SLM). An ultraviolet laser serves as a pump that illuminates a $\beta$-Barium Borate crystal (BBO). The output is filtered by LPF (a long pass filter) to remove residual pump. The SLM is placed in the Fourier plane of the crystal output surface. The second part comprises propagation in free space with a thin phase scatterer placed in the image plane of the SLM. The third part contains the detection system involving an Electron Multiplied Charge Coupled Device (EMCCD) camera to observe the far-field of the scattering medium and detect photon correlations. Lens $L_1$ is introduced to decrease the degree of entanglement and increase the spatial coherence of the SPDC beam, enabling the measurement of the transmission matrix (TM).}
\end{figure*}

The case of high-dimensional spatially-entangled two-photon states is particularly challenging. Unlike coherent classical light, these states do not produce high-contrast intensity speckle patterns after propagation through a complex medium. This can be understood through the analogy between a reduced single-photon state (that describes the intensity observed for a two-photon state) and partially coherent light~\cite{saleh2000duality}; as entanglement dimensionality increases, first-order coherence decreases, and interference effects in the intensity wash out. 
Conventional wavefront shaping, relying on first-order coherence and intensity feedback, is thus very difficult to implement under these conditions.
Nevertheless, these states do exhibit second-order coherence i.e. speckle patterns in their correlation measurements~\cite{beenakker_two-photon_2009,peeters_observation_2010,aarav2025new}, which can be leveraged for wavefront correction, as demonstrated in Refs.~\cite{valencia2020unscrambling,courme2025non}. However, a major limitation in this case is the prohibitively long acquisition time for photon correlation measurements - from several hours to days with current technologies, depending on the medium's complexity and the state dimensionality.

In this letter, we propose a protocol that uses the two-photon state itself to perform TM-based wavefront shaping through a scattering medium without the need for a classical beacon or correlation measurements. By harnessing the duality between partial coherence and entanglement, we tailor the entanglement of the source to reduce its dimensionality, generating two-photon states that regain first-order spatial coherence and exhibit speckle patterns directly in intensity images. These low-Schmidt-number states behave almost like spatially coherent classical light, enabling direct measurement of the medium's TM, which can then be used to control the propagation of high-dimensional entangled states. The same SLM used to measure the correction is used to shape the entangled state as well, providing a resource-minimal approach to quantum wavefront correction.

\begin{figure*}[t]
\centering
\includegraphics[keepaspectratio=true, scale = .74,trim = {.85cm 22.5cm 2.5cm .5cm},clip]{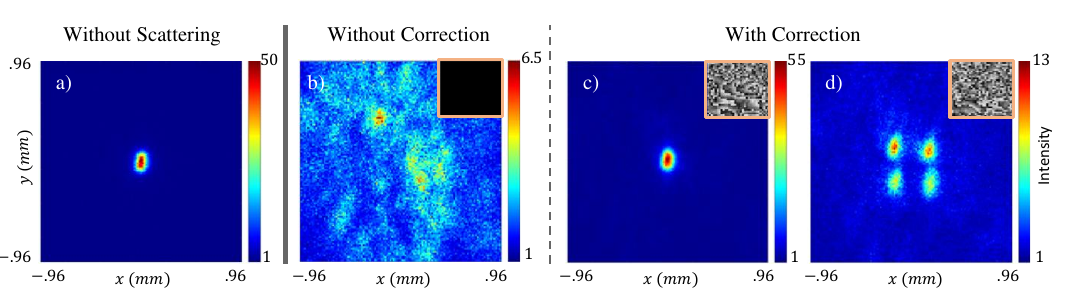}% Here is how to import EPS art
\caption{\label{fig_classical} \textbf{Wavefront correction for a low-Schmidt-number two-photon state.} (a) Intensity image measured without the scattering medium. (b) Intensity speckle pattern obtained in the presence of the medium.  (c) After wavefront correction, a focused spot appears in the intensity image. (d) Intensity image obtained using the SLM to simultaneously correct the wavefront and shape the beam into four foci.All images were acquired with EMCCD gain 200. Exposure times: 5 ms for (a), 50 ms for (b–d). Insets show the SLM patterns used with the scattering medium.}
\end{figure*}

\section{Concept}
A spatially-entangled two-photon state is generated via Spontaneous Parametric Down-Conversion (SPDC)~\cite{karan2020phase}. 
%As shown in Figure~\ref{fig_setup}, a pump laser illuminates a nonlinear crystal, resulting in the emission of a pair of entangled photons. 
Considering frequency- and polarization-degenerate photon pairs (type I SPDC) and assuming a Gaussian pump beam, the two-photon wavefunction at the output of the crystal can be modeled as a double-Gaussian distribution in transverse position space ~\cite{fedorov2009gaussian,schneeloch2016introduction}:
\begin{equation}
    \psi(\vec{r_i},\vec{r_s}) = A e^{-\frac{|\vec{r_i}+\vec{r_s}|^2 \sigma_\vec{p}^2}{4}}e^{-\frac{|\vec{r_i}-\vec{r_s}|^2}{4\sigma_\vec{r}^2}},
\end{equation}
where $A$ is a normalization parameter, $\vec{r_i}$ and $\vec{r_s}$ denote the transverse positions of idler and signal photons, respectively. $\sigma_\vec{r}=\sqrt{L \lambda_p / (6 \pi)}$ and $\sigma_\vec{p}=1/ (2 w)$ are the position and momentum correlation widths, with $L$ the crystal thickness, $\lambda_p$ the pump wavelength in the crystal, and $w$ the pump beam waist. When the pump beam diameter is sufficiently large (e.g. collimated beam), the system operates in the regime $1/\sigma_\vec{p} \gg \sigma_\vec{r}$. In this case, the state is entangled in many dimensions and exhibits low first-order spatial coherence, resulting in interference patterns with very low contrast in intensity. When the pump is focused on the crystal plane, $1/\sigma_\vec{p} \approx \sigma_r$ and the coherence properties of the two-photon state are significantly modified, approaching those of a classical coherent beam. These variations in coherence and entanglement dimensionality are quantified by the following expressions (adapted to two dimensions from \cite{schneeloch2016introduction}):
\begin{equation}
    g^{(1)}(\vec{r},-\vec{r}) = e^{-\frac{|\vec{r}|^2}{8w^2}\frac{(24\pi w^2 -L\lambda_p)^2}{(24\pi w^2 +L\lambda_p)L\lambda_p}}
\end{equation}
and
\begin{equation}
    K = \frac{1}{4} \left( \frac{24\pi w^2 +L\lambda_p}{\sqrt{24\pi w^2}\sqrt{L\lambda_p}}\right)^2,
\end{equation}
where $ g^{(1)}$ is the first-order spatial coherence function and $K$ the Schmidt number. 
$g^{(1)}(\vec{r},-\vec{r})$ characterizes the field-correlations between transverse positions $\vec{r}$ and $-\vec{r}$, with its width $\Delta g^{(1)}$ determining the spatial coherence length of the SPDC light, while $K$ quantifies the entanglement dimensionality of the generated state. 
As the pump waist $w$ decreases, $\Delta g^{(1)}$ increases while $K$ decreases, reflecting the transfer of entanglement to spatial coherence. In this work, we leverage this effect to measure efficiently the TM of a scattering medium and the associated correction.

\section{Experiment}
Figure~\ref{fig_setup} shows the experimental setup. A pump laser ($\lambda_p = 406 \text{nm},w \approx 1 \text{mm}$) illuminates a $1$-mm-thick $\beta$-Barium Borate (BBO) nonlinear crystal, resulting in the emission of entangled photon pairs.
The crystal output surface is Fourier-imaged onto a phase-only SLM (Holoeye) by lens $L_2$, which consists of three lenses ($5$ cm–$15$ cm–$15$ cm) arranged in a confocal configuration. 
The SLM is then imaged onto the scattering medium (a layer of parafilm) by a telescope formed by lenses $L_3$ ($15$ cm) and $L_4$ ($10$ cm). Finally, lens $L_5$ ($10$ cm) Fourier-images the output surface of the scattering medium onto an EMCCD camera. 
$LPF_1$ and $LPF_2$ are long-pass filters with cut-off wavelengths of $450$ nm and $750$ nm, respectively.

When the pump beam is collimated at the crystal plane (no lens $L_1$), the SPDC beam is entangled in high dimensions and exhibits very low spatial incoherence i.e. $K \approx 37$ and $\Delta g^{(1)} \approx \sigma_\vec{r} = 11 \mu m$. Introducing lens $L_1$ ($10\,\text{cm}$) focuses the pump onto the crystal plane, enabling the SPDC beam to regain spatial coherence while losing entanglement. 
In this configuration, a well-defined focal spot is observed on the camera in the absence of a scattering medium (Fig.~\ref{fig_classical}a), while the insertion of the medium produces a speckle intensity pattern (Fig.~\ref{fig_classical}b).
To perform wavefront correction, we use the TM-based method introduced in Ref.~\cite{popoff2010measuring}. 
First, we measure a $1024 \times 14641$ matrix with an $800$ ms exposure and EMCCD gain of $200$. The active SLM area is $200 \times 200$ pixels, divided into $32 \times 32$ macro-pixels. 
Then, we use the matrix to generate and display a phase pattern on the SLM that re-focuses light at the image center (Fig.~\ref{fig_classical}c), as if no scattering medium were present. With full control over the pupil plane, we can also shape the beam arbitrarily. For instance, in Figure~\ref{fig_classical}d, the same matrix is used to focus the beam at four different spots.

\begin{figure*}[t]
\centering
\includegraphics[keepaspectratio=true, scale = .74,trim = {1.15cm 18.2cm 2.5cm .9cm},clip]{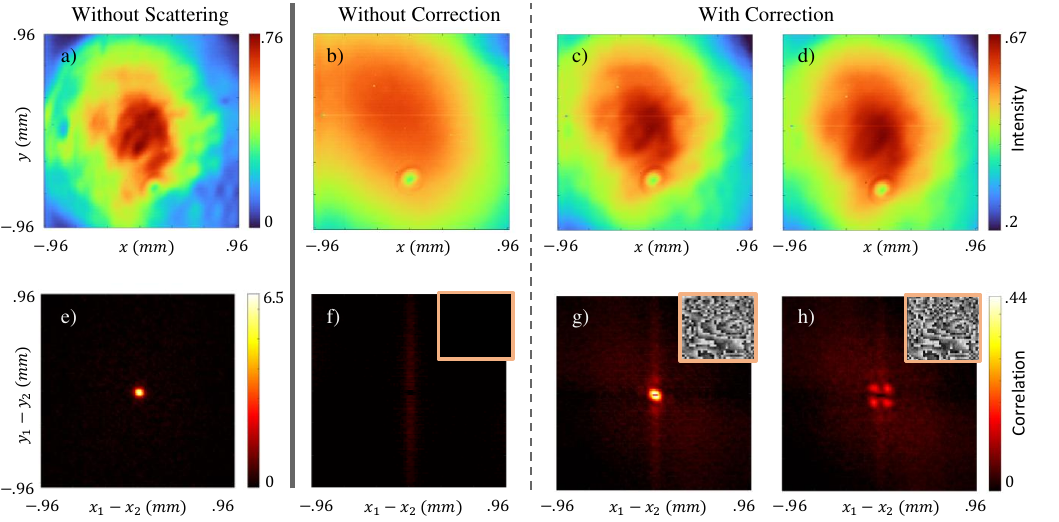}% Here is how to import EPS art
\caption{\label{fig_quantum} \textbf{Wavefront correction for a high-Schmidt-number two-photon state.} The top row depicts the intensity images, and the bottom row depicts the spatially-resolved second-order correlation function $G^{(2)}$ projected on the minus coordinates, for different experimental configurations. 
(a) Intensity image measured in the absence of a scattering medium, (b) with the scattering medium present and no correction applied (flat SLM), (c) after applying a SLM correction to focus classical light on a single point, and (d) after applying a SLM correction to focus correlations on four points.
(e) Sum-coordinate projections of $G^{(2)}$ measured in the absence of a scattering medium, (f) with the scattering medium present and no correction applied (flat SLM), (g) after applying a SLM correction to focus classical light on a single point, and (h) after applying a SLM correction to focus correlations on four points.
The SLM patterns used are shown in insets. For (d,h), it is generated by adding a $\pi$ phase shift to the top-left quadrant of the phase used for (c,g). Acquisition parameters - frames, exposure, EMCCD gain: (e) $3.4 \times 10^4$ frames, 1 ms, gain 200; (f) $2.4 \times 10^7$ frames, 3 ms, gain 1000; (g,h) $8.6 \times 10^6$ frames, 2 ms, gain 1000. The minus-coordinate projection images were denoised using a low-pass filter.}
\end{figure*}

Now that the TM of the medium is known, we use it to transmit a high-Schmidt-number entangled state. This is achieved by removing the lens $L_1$, so that the pump is not focused on the crystal anymore. 
{In this case, the output intensity is spatially homogeneous without any distinctive features (Fig.~\ref{fig_quantum}a), and retains this property in the presence of a scattering medium or when a phase mask is displayed on the SLM (Figs.\ref{fig_quantum}b–d).}
To characterize the transmitted output state, we measure the second-order correlation function $G^{(2)}$ with the EMCCD camera using the technique outlined in Ref.~\cite{defienne2018general}. Specifically, we focus on the minus-coordinate projection of $G^{(2)}$, which represents the joint probability of detecting two photons as a function of their relative distance on the camera.
As shown in Figure~\ref{fig_quantum}e, in the absence of a scattering medium, the entangled state produces a sharp central correlation peak in the minus-coordinate projection, reflecting strong multimode spatial correlations between photon pairs arising from high-dimensional entanglement~\cite{edgar_imaging_2012,moreau_realization_2012}. 

In our experiment, this correlation peak is used as an indicator of the quality of the applied correction. When no phase mask is displayed on the SLM, the correlation image does not show any peak; any speckles in the correlation are overwhelmed by detector noise (Fig.~\ref{fig_quantum}f). 
However, when we apply the same SLM pattern that was used to focus the classical beam (Fig.~\ref{fig_classical}b), a strong correlation peak is restored in the projection (Fig. \ref{fig_quantum}g). Additionally, subtle features in the corrected intensity image (Fig.~\ref{fig_quantum}c), such as smudges and dust on the crystal surface, become faintly visible, offering a weak classical signature of the correction.
In Figure~\ref{fig_quantum}h, we use the TM and SLM to simultaneously correct for scattering and shape in the spatial correlations i.e. the projection shows four spots instead of one, effectively encoding the information in correlations, as studied in previous works~\cite{verniere2024hiding,cameron2024shaping}. 

\section{Discussion}
\begin{figure*}[t]
\centering
\includegraphics[keepaspectratio=true, scale = .74,trim = {0cm 21.5cm 4cm 0.cm},clip]{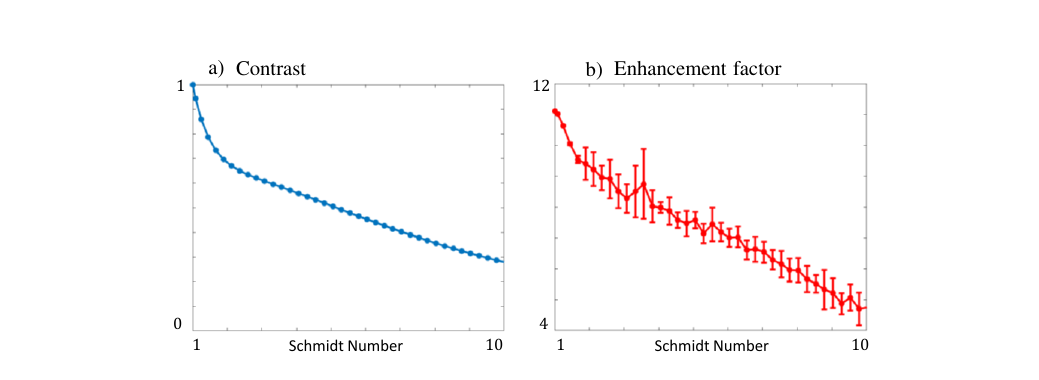}% Here is how to import EPS art
\caption{\label{fig_schmidt} \textbf{Simulations to study the effects of Schmidt number $K$ on wavefront correction.} (a) Normalized speckle contrast as a function of the state’s Schmidt number $K$. (b) Focusing enhancement factor (normalized) for a coherent state ($K = 1$) using the phase masks learned via states with varying Schmidt numbers $K$. The simulations were performed in one spatial dimension. The simulated beam extended over a region of size $10$mm, discretized into $1500$ pixels. The near field two-photon correlation width was fixed at $133\mu$m, and $K$ was varied by adjusting the pump beam width. The wavelength of light used was $810$nm.}
\end{figure*}

Finally, we investigate how the Schmidt number $K$ of a two-photon state affects wavefront correction performance using numerical simulations. The simulations employ the same imaging geometry as the experimental setup, with a random phase profile acting like a thin scattering medium. 
As shown in Figure~\ref{fig_schmidt}a, the speckle contrast decreases with increasing $K$. Here, the contrast is defined as the standard deviation divided by the mean intensity in the central speckle region, normalized to the maximum observed value. 
This reduction in contrast reflects the loss of spatial coherence in the beam and is expected to reduce the efficiency of wavefront optimization. 

To observe this effect, we study the optimization efficiency as a function of $K$ in Figure~\ref{fig_schmidt}b. For each two-photon beam, we optimized the intensity at a target pixel, using the method outlined in Ref.~\cite{vellekoop2007focusing}. Each optimization was run for $250$ iterations to generate a phase mask. 
To benchmark the efficacy of each mask, we propagated a coherent beam ($K=1$) through the same scatterer plus the phase mask, recorded the resulting target intensity, and calculated the enhancement factor ($\eta$). 
Here, the enhancement factor is the ratio of the target intensity after optimization to the average speckle intensity before optimization. This process was repeated $10$ times for each value of $K$, and the mean and standard deviation of $\eta$ are plotted in Figure~\ref{fig_schmidt}b. It shows that, just like the intensity contrast goes down with $K$, the efficiency factor goes down with increasing $K$ as well, indicating the failure of the optimization process in the high-entanglement regime. 

\section{Conclusion}
In this work, we demonstrated a fast and robust method to restore spatial correlations of a high-dimensional two-photon state through a scattering medium without the need for any classical beacon beam. Our approach harnesses the fundamental duality between coherence and entanglement in two-photon states and requires no additional components beyond those already used for shaping entangled photon pairs. It involves two steps: first, by focusing the pump beam, we generate a low-Schmidt-number two-photon state that exhibits speckle patterns in intensity, mimicking classical coherent light, and we use the output intensity to retrieve the TM with an SLM. Second, the same SLM is used both to correct the wavefront distortion and to structure the high-dimensional two-photon entangled state. 

In the present case, we used a single SLM to perform the correction, which is well-suited for a thin phase-scattering medium. However, if the scatterer is thick or is not in the image plane of the SLM, the correction cannot be done optimally, and other alternatives, such as multi-plane light converters (MPLC) \cite{kupianskyi2024all}, need to be employed. Another limitation arises from the low intensity of the SPDC light compared to bright lasers, making this method slower, which could be problematic in cases of dynamic scatterers. 

Importantly, because this technique relies solely on the SPDC-generated two-photon state itself, it inherently matches the intended quantum state in key characteristics such as polarization, temporal bandwidth, and wavelength. For example, spectral mode matching is essential in complex disordered media such as multimode fibers or thick scattering materials, which are subject to modal dispersion. In such systems, efficient light shaping is only possible if the TM and the controlled light share the same spectral mode~\cite{mounaix_spatiotemporal_2016,small_spectral_2012,beijnum_frequency_2011,mounaix_spatiotemporal_2016}. In addition, our approach is especially well-suited for scenarios where generating or integrating a classical reference is impractical or impossible. For example, Ghost imaging involving different wavelengths~\cite{aspden_photon-sparse_2015} would require a separate classical beacon at each wavelength. We anticipate that this protocol will be valuable in advancing quantum communication and imaging through complex media, especially in high-dimensional or resource-constrained settings.

\begin{backmatter}
\bmsection{Funding}
H.D. acknowledges funding from the ERC Starting Grant (No. SQIMIC-101039375).

\bmsection{Disclosures}
The authors declare no conflicts of interest.

\bmsection{Data Availability Statement}
Data underlying the results presented in this paper are not publicly available at this time but may be obtained from the authors upon reasonable request.

\end{backmatter}

%%%%%%%%%% If using BibTeX:
\bibliography{Ref}

\begin{thebibliography}{10}
\newcommand{\enquote}[1]{``#1''}

\bibitem{erhard_advances_2020}
M.~Erhard, M.~Krenn, and A.~Zeilinger, \enquote{Advances in high-dimensional quantum entanglement,} {\protect\JournalTitle{Nature Reviews Physics}} \textbf{2}, 365--381 (2020). Publisher: Nature Publishing Group UK London.

\bibitem{walborn_transverse_2007}
S.~P. Walborn and C.~H. Monken, \enquote{Transverse spatial entanglement in parametric down-conversion,} {\protect\JournalTitle{Physical Review A}} \textbf{76}, 062305 (2007).

\bibitem{defienne_advances_2024}
H.~Defienne, W.~P. Bowen, M.~Chekhova, \emph{et~al.}, \enquote{Advances in quantum imaging,} {\protect\JournalTitle{Nature Photonics}} \textbf{18}, 1024--1036 (2024). Publisher: Nature Publishing Group.

\bibitem{brida2010experimental}
G.~Brida, M.~Genovese, and I.~Ruo~Berchera, \enquote{Experimental realization of sub-shot-noise quantum imaging,} {\protect\JournalTitle{Nature Photonics}} \textbf{4}, 227--230 (2010).

\bibitem{defienne_quantum_2019}
H.~Defienne, M.~Reichert, J.~W. Fleischer, and D.~Faccio, \enquote{Quantum image distillation,} {\protect\JournalTitle{Science Advances}} \textbf{5}, eaax0307 (2019).

\bibitem{gregory_noise_2021}
T.~Gregory, P.-A. Moreau, S.~Mekhail, \emph{et~al.}, \enquote{Noise rejection through an improved quantum illumination protocol,} {\protect\JournalTitle{Scientific reports}} \textbf{11}, 21841 (2021). Publisher: Nature Publishing Group UK London.

\bibitem{camphausen_quantum-enhanced_2021}
R.~Camphausen, A.~Cuevas, L.~Duempelmann, \emph{et~al.}, \enquote{A quantum-enhanced wide-field phase imager,} {\protect\JournalTitle{Science Advances}} \textbf{7}, eabj2155 (2021).

\bibitem{toninelli_resolution-enhanced_2019}
E.~Toninelli, P.-A. Moreau, T.~Gregory, \emph{et~al.}, \enquote{Resolution-enhanced quantum imaging by centroid estimation of biphotons,} {\protect\JournalTitle{Optica}} \textbf{6}, 347--353 (2019). Publisher: Optical Society of America.

\bibitem{defienne2022pixel}
H.~Defienne, P.~Cameron, B.~Ndagano, \emph{et~al.}, \enquote{Pixel super-resolution with spatially entangled photons,} {\protect\JournalTitle{Nature communications}} \textbf{13}, 3566 (2022).

\bibitem{he_quantum_2023}
Z.~He, Y.~Zhang, X.~Tong, \emph{et~al.}, \enquote{Quantum microscopy of cells at the {Heisenberg} limit,} {\protect\JournalTitle{Nature Communications}} \textbf{14}, 2441 (2023). Number: 1 Publisher: Nature Publishing Group.

\bibitem{cameron2024adaptive}
P.~Cameron, B.~Courme, C.~Verni{\`e}re, \emph{et~al.}, \enquote{Adaptive optical imaging with entangled photons,} {\protect\JournalTitle{Science}} \textbf{383}, 1142--1148 (2024).

\bibitem{srivastav_quick_2022}
V.~Srivastav, N.~H. Valencia, W.~McCutcheon, \emph{et~al.}, \enquote{Quick {Quantum} {Steering}: {Overcoming} {Loss} and {Noise} with {Qudits},} {\protect\JournalTitle{Physical Review X}} \textbf{12}, 041023 (2022).

\bibitem{ecker2019overcoming}
S.~Ecker, F.~Bouchard, L.~Bulla, \emph{et~al.}, \enquote{Overcoming noise in entanglement distribution,} {\protect\JournalTitle{Physical Review X}} \textbf{9}, 041042 (2019).

\bibitem{zhu_is_2021}
F.~Zhu, M.~Tyler, N.~H. Valencia, \emph{et~al.}, \enquote{Is high-dimensional photonic entanglement robust to noise?} {\protect\JournalTitle{AVS Quantum Science}} \textbf{3}, 011401 (2021).

\bibitem{cozzolino_high-dimensional_2019}
D.~Cozzolino, B.~Da~Lio, D.~Bacco, and L.~K. Oxenløwe, \enquote{High-{Dimensional} {Quantum} {Communication}: {Benefits}, {Progress}, and {Future} {Challenges},} {\protect\JournalTitle{Advanced Quantum Technologies}} \textbf{2}, 1900038 (2019).

\bibitem{defienne_two-photon_2016}
H.~Defienne, M.~Barbieri, I.~A. Walmsley, \emph{et~al.}, \enquote{Two-photon quantum walk in a multimode fiber,} {\protect\JournalTitle{Science Advances}} \textbf{2}, e1501054 (2016).

\bibitem{defienne2018adaptive}
H.~Defienne, M.~Reichert, and J.~W. Fleischer, \enquote{Adaptive quantum optics with spatially entangled photon pairs,} {\protect\JournalTitle{Physical review letters}} \textbf{121}, 233601 (2018).

\bibitem{courme_manipulation_2023}
B.~Courme, P.~Cameron, D.~Faccio, \emph{et~al.}, \enquote{Manipulation and {Certification} of {High}-{Dimensional} {Entanglement} through a {Scattering} {Medium},} {\protect\JournalTitle{PRX Quantum}} \textbf{4}, 010308 (2023). Publisher: American Physical Society.

\bibitem{devaux_restoring_2022}
F.~Devaux, A.~Mosset, S.~M. Popoff, and E.~Lantz, \enquote{Restoring and tailoring very high dimensional spatial entanglement of a biphoton state transmitted through a scattering medium,}  (2022). ArXiv:2206.00299 [physics, physics:quant-ph].

\bibitem{cao_shaping_2022}
H.~Cao, A.~P. Mosk, and S.~Rotter, \enquote{Shaping the propagation of light in complex media,} {\protect\JournalTitle{Nature Physics}} \textbf{18}, 994--1007 (2022). Publisher: Nature Publishing Group.

\bibitem{xu2018imaging}
X.~Xu, X.~Xie, A.~Thendiyammal, \emph{et~al.}, \enquote{Imaging of objects through a thin scattering layer using a spectrally and spatially separated reference,} {\protect\JournalTitle{Optics express}} \textbf{26}, 15073--15083 (2018).

\bibitem{mounaix_spatiotemporal_2016}
M.~Mounaix, D.~Andreoli, H.~Defienne, \emph{et~al.}, \enquote{Spatiotemporal {Coherent} {Control} of {Light} through a {Multiple} {Scattering} {Medium} with the {Multispectral} {Transmission} {Matrix},} {\protect\JournalTitle{Physical Review Letters}} \textbf{116}, 253901 (2016).

\bibitem{mounaix_control_2019}
M.~Mounaix and J.~Carpenter, \enquote{Control of the temporal and polarization response of a multimode fiber,} {\protect\JournalTitle{Nature Communications}} \textbf{10}, 1--8 (2019). Number: 1 Publisher: Nature Publishing Group.

\bibitem{aguiar_polarization_2017}
H.~B.~d. Aguiar, S.~Gigan, and S.~Brasselet, \enquote{Polarization recovery through scattering media,} {\protect\JournalTitle{Science Advances}} \textbf{3}, e1600743 (2017).

\bibitem{katz_focusing_2011}
O.~Katz, E.~Small, Y.~Bromberg, and Y.~Silberberg, \enquote{Focusing and compression of ultrashort pulses through scattering media,} {\protect\JournalTitle{Nature Photonics}} \textbf{5}, 372--377 (2011).

\bibitem{paudel2013focusing}
H.~P. Paudel, C.~Stockbridge, J.~Mertz, and T.~Bifano, \enquote{Focusing polychromatic light through strongly scattering media,} {\protect\JournalTitle{Optics express}} \textbf{21}, 17299--17308 (2013).

\bibitem{mounaix_deterministic_2016}
M.~Mounaix, H.~Defienne, and S.~Gigan, \enquote{Deterministic light focusing in space and time through multiple scattering media with a time-resolved transmission matrix approach,} {\protect\JournalTitle{Physical Review A}} \textbf{94}, 041802 (2016).

\bibitem{katz_looking_2012}
O.~Katz, E.~Small, and Y.~Silberberg, \enquote{Looking around corners and through thin turbid layers in real time with scattered incoherent light,} {\protect\JournalTitle{Nature Photonics}} \textbf{6}, 549--553 (2012). Number: 8 Publisher: Nature Publishing Group.

\bibitem{aspden_photon-sparse_2015}
R.~S. Aspden, N.~R. Gemmell, P.~A. Morris, \emph{et~al.}, \enquote{Photon-sparse microscopy: visible light imaging using infrared illumination,} {\protect\JournalTitle{Optica}} \textbf{2}, 1049--1052 (2015). Publisher: Optical Society of America.

\bibitem{lemos_quantum_2022}
G.~B. Lemos, M.~Lahiri, S.~Ramelow, \emph{et~al.}, \enquote{Quantum imaging and metrology with undetected photons: tutorial,} {\protect\JournalTitle{JOSA B}} \textbf{39}, 2200--2228 (2022). Publisher: Optica Publishing Group.

\bibitem{saleh2000duality}
B.~E. Saleh, A.~F. Abouraddy, A.~V. Sergienko, and M.~C. Teich, \enquote{Duality between partial coherence and partial entanglement,} {\protect\JournalTitle{Physical Review A}} \textbf{62}, 043816 (2000).

\bibitem{beenakker_two-photon_2009}
C.~W.~J. Beenakker, J.~W.~F. Venderbos, and M.~P. van Exter, \enquote{Two-{Photon} {Speckle} as a {Probe} of {Multi}-{Dimensional} {Entanglement},} {\protect\JournalTitle{Physical Review Letters}} \textbf{102}, 193601 (2009).

\bibitem{peeters_observation_2010}
W.~H. Peeters, J.~J.~D. Moerman, and M.~P. van Exter, \enquote{Observation of {Two}-{Photon} {Speckle} {Patterns},} {\protect\JournalTitle{Physical Review Letters}} \textbf{104} (2010).

\bibitem{aarav2025new}
S.~Aarav, S.~Wadood, and J.~W. Fleischer, \enquote{New scattering zones in quantum speckle propagation,} {\protect\JournalTitle{arXiv preprint arXiv:2507.08408}}  (2025).

\bibitem{valencia2020unscrambling}
N.~H. Valencia, S.~Goel, W.~McCutcheon, \emph{et~al.}, \enquote{Unscrambling entanglement through a complex medium,} {\protect\JournalTitle{Nature Physics}} \textbf{16}, 1112--1116 (2020).

\bibitem{courme2025non}
B.~Courme, C.~Verni{\`e}re, M.~Joly, \emph{et~al.}, \enquote{Non-classical optimization through complex media,} {\protect\JournalTitle{arXiv preprint arXiv:2503.24283}}  (2025).

\bibitem{karan2020phase}
S.~Karan, S.~Aarav, H.~Bharadhwaj, \emph{et~al.}, \enquote{Phase matching in $\beta$-barium borate crystals for spontaneous parametric down-conversion,} {\protect\JournalTitle{Journal of Optics}} \textbf{22}, 083501 (2020).

\bibitem{fedorov2009gaussian}
M.~Fedorov, Y.~M. Mikhailova, and P.~Volkov, \enquote{Gaussian modelling and schmidt modes of spdc biphoton states,} {\protect\JournalTitle{Journal of Physics B: Atomic, Molecular and Optical Physics}} \textbf{42}, 175503 (2009).

\bibitem{schneeloch2016introduction}
J.~Schneeloch and J.~C. Howell, \enquote{Introduction to the transverse spatial correlations in spontaneous parametric down-conversion through the biphoton birth zone,} {\protect\JournalTitle{Journal of Optics}} \textbf{18}, 053501 (2016).

\bibitem{popoff2010measuring}
S.~M. Popoff, G.~Lerosey, R.~Carminati, \emph{et~al.}, \enquote{Measuring the transmission matrix in optics: An approach to the study and control of light propagation in disordered media,} {\protect\JournalTitle{Physical review letters}} \textbf{104}, 100601 (2010).

\bibitem{defienne2018general}
H.~Defienne, M.~Reichert, and J.~W. Fleischer, \enquote{General model of photon-pair detection with an image sensor,} {\protect\JournalTitle{Physical review letters}} \textbf{120}, 203604 (2018).

\bibitem{edgar_imaging_2012}
M.~P. Edgar, D.~S. Tasca, F.~Izdebski, \emph{et~al.}, \enquote{Imaging high-dimensional spatial entanglement with a camera,} {\protect\JournalTitle{Nature Communications}} \textbf{3}, 984 (2012). Number: 1 Publisher: Nature Publishing Group.

\bibitem{moreau_realization_2012}
P.-A. Moreau, J.~Mougin-Sisini, F.~Devaux, and E.~Lantz, \enquote{Realization of the purely spatial {Einstein}-{Podolsky}-{Rosen} paradox in full-field images of spontaneous parametric down-conversion,} {\protect\JournalTitle{Physical Review A}} \textbf{86}, 010101 (2012).

\bibitem{verniere2024hiding}
C.~Verni{\`e}re and H.~Defienne, \enquote{Hiding images in quantum correlations,} {\protect\JournalTitle{Physical Review Letters}} \textbf{133}, 093601 (2024).

\bibitem{cameron2024shaping}
P.~Cameron, B.~Courme, D.~Faccio, and H.~Defienne, \enquote{Shaping the spatial correlations of entangled photon pairs,} {\protect\JournalTitle{Journal of Physics: Photonics}} \textbf{6}, 033001 (2024).

\bibitem{vellekoop2007focusing}
I.~M. Vellekoop and A.~P. Mosk, \enquote{Focusing coherent light through opaque strongly scattering media,} {\protect\JournalTitle{Optics letters}} \textbf{32}, 2309--2311 (2007).

\bibitem{kupianskyi2024all}
H.~Kupianskyi, S.~A. Horsley, and D.~B. Phillips, \enquote{All-optically untangling light propagation through multimode fibers,} {\protect\JournalTitle{Optica}} \textbf{11}, 101--112 (2024).

\bibitem{small_spectral_2012}
E.~Small, O.~Katz, Y.~Guan, and Y.~Silberberg, \enquote{Spectral control of broadband light through random media by wavefront shaping,} {\protect\JournalTitle{Optics Letters}} \textbf{37}, 3429 (2012).

\bibitem{beijnum_frequency_2011}
F.~v. Beijnum, E.~G.~v. Putten, A.~Lagendijk, and A.~P. Mosk, \enquote{Frequency bandwidth of light focused through turbid media,} {\protect\JournalTitle{Optics Letters}} \textbf{36}, 373--375 (2011).

\end{thebibliography}

\end{document}